\documentclass[conference]{IEEEtran}
\IEEEoverridecommandlockouts
\usepackage{cite}
\usepackage{amsmath,amssymb,amsfonts}
\usepackage{algorithmic}
\usepackage{graphicx}
\usepackage{pgf-pie} 
\usepackage{textcomp}
\usepackage{xcolor}
\usepackage{caption}
\usepackage{subcaption}
\usepackage{makecell, cellspace}
\usepackage{hhline}
\usepackage{color, colortbl}
\usepackage{multirow}
\usepackage{tcolorbox}

\usepackage{algorithm}
\usepackage{caption}

\def\BibTeX{{\rm B\kern-.05em{\sc i\kern-.025em b}\kern-.08em
    T\kern-.1667em\lower.7ex\hbox{E}\kern-.125emX}}
\begin{document}

\title{FlaKat: A Machine Learning-Based\\ 
Categorization Framework for Flaky Tests
}
\author{
    \IEEEauthorblockN{Shizhe Lin, Ryan Liu, Ladan Tahvildari}
    \IEEEauthorblockA{
        \textit{Dept. of Electrical and Computer Engineering} \\
        \textit{University of Waterloo, Canada}\\
        \{s222lin, ryan.zheng.he.liu, ladan.tahvildari\}@uwaterloo.ca
    }
}    
\maketitle

\begin{abstract}

Flaky tests can pass or fail non-deterministically, without alterations to a software system. Such tests are frequently encountered by developers and hinder the credibility of test suites. State-of-the-art research incorporates machine learning solutions into flaky test detection and achieves reasonably good accuracy. Moreover, the majority of automated flaky test repair solutions are designed for specific types of flaky tests. This research work proposes a novel categorization framework, called FlaKat, which uses machine-learning classifiers for fast and accurate prediction of the category of a given flaky test that reflects its root cause. Sampling techniques are applied to address the imbalance between flaky test categories in the International Dataset of Flaky Test (IDoFT). A new evaluation metric, called Flakiness Detection Capacity (FDC), is proposed for measuring the accuracy of classifiers from the perspective of information theory and provides proof for its effectiveness. The final FDC results are also in agreement with \(F_1\) score regarding which classifier yields the best flakiness classification.
\end{abstract}


\section{Introduction}
\label{sec:introduction}

Flaky tests that pass and fail arbitrarily with the same software-under-test are common phenomena in modern-day software development. They cause unnecessary wasted efforts during continuous integration and regression testing, which are widely adopted in both industry and the open-source community~\cite{ref:whenWeTalkFlakiness}. Additionally, failures due to test case flakiness also hinders the reliability and validity of methodologies proposed by researchers, especially in areas such as fault localization~\cite{ref:faultLocalization} and mutation testing~\cite{ref:mutationTesting}. The fact that flaky tests are frequently encountered makes the situation worse with 59\% of respondents from a survey of industrial developers claiming to experience it at least monthly. Moreover, 91\% of the developers claimed to deal with flaky tests at least a few times in a year~\cite{ref:DeveloperPerspective}.


Researchers have studied flaky tests since 2009 and developed numerous solutions for flaky test detection~\cite{ref:killingGatekeeper}. The traditional approaches usually adopt dynamic analysis that executes test suites repeatedly~\cite{ref:largeScaleStudy}. For certain flaky tests, it may require hundreds, sometimes thousands, of executions for the failure to occur, which makes them more troublesome to identify~\cite{ref:flakeFlagger,ref:modelRankAtApple}. Some other flaky tests may only pass or fail if the test suites are executed in a specific order, which requires formulating randomized execution plans and collecting the results from all attempts~\cite{ref:eventOrder,ref:iDflakies}. Despite being effective in detecting flaky tests, these tools are also relatively expensive in terms of computational power and runtime. Some flakiness detection tools take the static analysis approach~\cite{ref:assosiationRule,ref:BayesianNetwork} and recent studies also incorporated machine learning tools into detecting test case flakiness with promising results~\cite{ref:vocabOfFlaky}. Compared to the traditional methods, machine learning solutions are much faster at locating flaky tests in the test suites.

Once identified, flaky tests can be repaired according to their categories and root causes manually~\cite{ref:DeveloperPerspective,ref:empiricalAnalysisFlakyTest2014}. Recent studies also developed automated solutions to repair flaky tests for certain categories of flaky tests~\cite{ref:iFixFlakies,ref:DexFix} but there exists a gap between the tools for flaky test detection and category-specific flaky test fixes. Motivated by the prior works applying machine learning models in the area of flaky test detection~\cite{ref:flakeFlagger} and~\cite{ref:FLAST}, we propose a novel framework named FlaKat that adopts the machine learning approach to predict the \textit{category} of flaky tests. Our work explores the effectiveness of three different source code vectorization methods: doc2vec~\cite{ref:doc2vec}, code2vec~\cite{ref:code2vec}, and term frequency-inverse document frequency (tf-idf)~\cite{ref:tfidf}. Several dimensionality reduction techniques are experimented and the machine learning classifier used in the framework is also selected after careful evaluation and comparison in terms of \(F_1\) score and a new evaluation metric named Flakiness Detection Capacity \textit{(FDC)}. \(F_1\) score is an effective and popular indicator of model quality since it reflects both precision and recall~\cite{ref:improvedSoftwareCategorization}. FDC is derived from intrusion detection capacity for intrusion detection systems~\cite{ref:intrusionDetection} and can measure the performance of a classifier from the perspective of information-theoretic analysis. The development of FlaKat also served as the foundation of the Master thesis of the author~\cite{ref:thesis}.

The rest of this paper is divided into the following sections. Section~\ref{sec:methodology} describes the methodology and implementation details behind the proposed FlaKat framework as well as the research questions answered. Section~\ref{sec:evaluation} illustrates the evaluation of the framework via two metrics, \(F_1\) score and FDC. Section~\ref{sec:relatedWork} provides related work, including both studies on flaky tests and their state-of-the-art solutions. Section~\ref{sec:threatToValidity} goes through the threats to the validity of this research. Section~\ref{sec:conclusion} concludes the paper as well as points out some future directions.

\section{FlaKat Framework}
\label{sec:methodology}

\subsection{Motivation}
\label{sec:motivation}

Incorporating machine learning algorithms into the identification of flaky tests is an effective solution according to recent studies~\cite{ref:flakeFlagger,ref:FLAST,ref:vocabOfFlaky}. One approach is to vectorize test case source code using the basic bag-of-words technique. Then, use machine learning classifiers, such as Random Forest, to predict whether a given test case is flaky or not~\cite{ref:FLAST,ref:vocabOfFlaky}. Other approaches consider more factors related to the behaviour of test cases and extract additional source code features, such as the number of lines and test execution time~\cite{ref:flakeFlagger}. Both approaches can achieve flakiness predictions with good performance and low overhead. The most recent work tracking the category of flaky test fixes is FlakyCat~\cite{ref:flakyCat} where they classify flaky tests by keywords on embeddings vectorized by pre-trained neural network and achieved a weighted \(F_1\) of 0.7. They look at flakiness from a different perspective than IDoFT and could be a great aid in understanding the variation of flaky tests.

Although the state-of-the-art does not include a universal repair strategy for all types of flaky tests, there are plenty of flaky test repair techniques for addressing specific types of flaky tests. For instance, concurrency-related flakiness can be mitigated by adding \texttt{waitFor} methods for the asynchronous wait blocks~\cite{ref:DeveloperPerspective,ref:empiricalAnalysisFlakyTest2014}. Automatic tools have also been developed for order-dependent and implementation-dependent flaky tests~\cite{ref:iFixFlakies,ref:DexFix} and achieved successful fixes in real-life software projects to improve the reliability of their test suites. Recently, a deep-learning framework, FlakyFix~\cite{ref:flakyFix}, took advantage of the existing large language models and achieved satisfying recommendations of flaky test fixes based on the history of manual flaky test repairs.

So far, no framework effectively categorizes known flaky tests and labels them according to their behaviour and root cause. A potential approach to implementing this novel framework can adopt a machine-learning algorithm for fast and accurate classification while avoiding the runtime cost of repeated test execution. The new framework can provide categorical classifications following the standards established in prior studies, such as the distinction between order-dependent and implementation-dependent flaky tests~\cite{ref:DeveloperPerspective,ref:empiricalAnalysisFlakyTest2014}. Compared to the existing tools~\cite{ref:FLAST,ref:flakeFlagger}, developers can gain more insight into the nature of the flaky tests, avoid similar mistakes, and boost their productivity.

\subsection{Workflow and Implementation}
\label{sec:workflowAndImplementation}

The goal of our proposed framework, FlaKat, is to make predictions on the category of \textit{known} flaky tests given their raw source code (only limited to Java currently) and its workflow is shown in Alg.~\ref{alg:workflowCode}. The input to the framework is a \textit{known} flaky test from a collection of open-source projects on Github. The output of the framework is the categorical label for each given flaky test.

\newcommand{\INDSTATE}[1][1]{\STATE\hspace{#1\algorithmicindent}}

\begin{algorithm}[hbt!]
\caption{Workflow of FlaKat}\label{alg:workflowCode}
\begin{algorithmic}
\STATE \textbf{Input}: known flaky tests
\STATE \textbf{Output}: category of flaky tests
\STATE $rawData \gets knownFlakyTests$

\STATE $vectorEmbeddings \gets$ vectorize $rawData$

\STATE $reducedEmbeddings \gets$ dimensionality reduction on \\ 
\INDSTATE[11] $vectorEmbeddings$

\FORALL{$classifierHyperparameter$}
    \STATE $classifier \gets \text{create using} \ classifierHyperparameter$
    \FOR{$trainSet, testSet \gets split(reducedEmbeddings)$}
        \STATE $sampledSet \gets Sampling(trainSet)$
        \STATE $result \gets trainAndPredict(sampledSet,testSet)$
    \ENDFOR
\ENDFOR
\end{algorithmic}
\end{algorithm}

The FlaKat framework is implemented in Python available online\footnote{https://anonymous.4open.science/r/flakat-8C84} and makes use of existing machine learning packages, such as scikit-learn\footnote{https://scikit-learn.org/stable/}, imbalanced-learn\footnote{https://imbalanced-learn.org/stable/index.html} and gensim\footnote{https://radimrehurek.com/gensim/}. The FlaKat framework has the following four phases in its workflow:

\noindent{$\bullet$ \textbf{{Phase I-Parsing and Preprocessing:}}}
With the knowledge of where flaky tests are located in each project, irrelevant code was pruned so only the flaky test case source code was preserved. The raw test case source code was then flattened into a string and stored in a local file. A script developed for automatic downloading and pruning was used for extracting the source code of the targeted flaky test cases. With the correct project directory URL and the commit ID, the GitPython\footnote{https://gitpython.readthedocs.io/en/stable/} module can easily download the intended Java file from the project repository. Then, the contents of the files were tokenized using the Javalang\footnote{https://github.com/c2nes/javalang} module. The test cases were then parsed and stored in csv format. 

\noindent{$\bullet$ \textbf{{{Phase II-Vector Embedding Generation:}}}} Depending on the type of the vectorization technique, the raw test case string was converted into a vector of varying real numbers that can be processed by machine learning models. Various algorithms are available for this purpose and three were considered in our study. The first was tf-idf, which is a popular bag-of-words method for vectorizing documents and has been used in FLAST~\cite{ref:FLAST}. An alternative algorithm for text vectorization was the doc2vec, which is a generalization of the popular distributed semantic representation method word2vec~\cite{ref:doc2vec}. It captures the relationship between individual Java test cases in the test suites and examines flakiness from a different perspective from tf-idf. The code2vec~\cite{ref:code2vec} was the third source code vectorization technique used to generate embeddings. Unlike prior algorithms, the code2vec method views source code as Abstract Syntax Tree (AST) and preserves the structural information of the test case source code. 

\noindent{$\bullet$ \textbf{{Phase III-Dimension Reduction:}}}
Once the embeddings of raw test cases were generated, they need to be reduced to a lower dimension to limit training time. In addition, data points reduced to lower dimensions can also be visualized to understand their distribution and clustering. Popular dimensionality reduction techniques such as Principal Component Analysis (PCA) and Linear Discriminant Analysis (LDA) were applied. Isomap was also used to verify whether flaky test data points in high dimensions can be unfolded into lower dimensions. In addition, two manifold learning techniques, the t-distributed Stochastic Neighbor Embedding (t-SNE) as well as Uniform Manifold Approximation and Projection (UMAP)~\cite{ref:umap}, were applied and yielded interesting results here. The result of dimensionality reduction was analyzed both qualitatively and quantitatively. The optimal reduced embedding was then used as the input for training and testing in the next step.

\noindent{$\bullet$ \textbf{{{Phase IV-Sampling and Prediction Using Classifiers:}}}}
The sample size of different categories of flaky tests in the International Dataset of Flaky Test (IDoFT) dataset varies excessively, where the ratio between majority and minority could be as great as 50:1. Synthetic Minority Oversampling Technique (SMOTE) was used to oversample the rare flaky test categories and Tomek Link (TK) technique was applied to undersample the abundant ones. These techniques were applied on the training dataset throughout the process of cross-validation. Several popular classifiers from the scikit-learn library were selected. First, K-Nearest Neighbour (KNN) was the classifier adopted in FLAST~\cite{ref:FLAST} and has been shown to have high performance. Another algorithm with outstanding performance was Random Forest, which has been shown to surpass KNN and other models in flaky/non-flaky prediction~\cite{ref:vocabOfFlaky}. Bayesian Optimization was added for optimal tuning of the Random Forest classifier. Such method is supported by open-source project Bayesian Optimization \footnote{https://github.com/fmfn/BayesianOptimization}. Lastly, Support Vector Machine (SVM) was also considered since it is well-known for the non-linear classification of natural data point clusters. Two metrics were used to measure the performance of prediction, \(F_1\) score and Flakiness Detection Capacity (FDC). 

\subsection{Research Objectives}
\label{sec:problemStatement}

In this research, we present the FlaKat framework for fast and accurate labeling of the root cause of flaky tests via a machine-learning approach. During its realization, several research questions were to be answered.

\begin{tcolorbox}[width=\linewidth, sharp corners=all, colback=white!95!black]

\textbf{RQ1.} Is there a clean separation between clusters for different types of flaky tests in the vector space?

\end{tcolorbox}

The quality of data provided to the machine learning models has a crucial impact on their performance. After applying code vectorization, raw test cases were converted into arrays of real numbers in vector space. If data points from the same flaky category do not form distinct clusters in vector space, then accurate classification becomes impossible. Reducing the dimensionality of these arrays helps to visualize the data points and the two-dimensional scatter plots of vectorized flaky tests can provide direct indications regarding the degree of clustering. 

\begin{tcolorbox}[width=\linewidth, sharp corners=all, colback=white!95!black]

\textbf{RQ2.} Which dimension reduction technique best preserves the local and global structure of the flaky tests from the original higher-dimensional vector space?

\end{tcolorbox}

Different dimension reduction methods yield distinct results and strongly affect the performance of predictions later. Several popular reduction algorithms were applied. To evaluate dimensionality reduction algorithms used in FlaKat, direct inspection of reduced vector embedding is not accurate enough to draw a sound conclusion on which technique was most suitable for flaky test categorization. Using KNN algorithm added more confidence to the selection where results based on a small k value represent the preservation of local structure and results with a large k value represent the preservation of the global structure.

\begin{tcolorbox}[width=\linewidth, sharp corners=all, colback=white!95!black]

\textbf{RQ3.} Which classifier provides the most accurate predictions on the flakiness category measured in \(F_1\) score?

\end{tcolorbox}

\(F_1\) score is a widely-used metric for evaluating machine learning models~\cite{ref:improvedSoftwareCategorization}. It provides a quantitative value on how well the classifier predicts with data points from the test set. The optimal configuration of hyperparameters for the classifiers need to be explored and their performances compared. 

\begin{tcolorbox}[width=\linewidth, sharp corners=all, colback=white!95!black]

\textbf{RQ4.} Compared to F1 score, is Flaky Detection Capacity (FDC) a more consistent and discriminant metric for evaluating classification of flaky tests?

\end{tcolorbox}

FDC shown in Eq.\ref{equ:fdc} in Appendix~\ref{app:fdc} is the new metric inspired from earlier work on intrusion detection~\cite{ref:intrusionDetection}, which focused on the correlation between input and output of classifiers based on information theory. It is the ratio between the mutual information of vectorized embedding input and its category output to the entropy of the input. With limited knowledge of the importance of the different categories of flaky tests and their weights in calculating average \(F_1\) score, FDC can potentially provide more valuable measurement on the performance of classifiers compared to calculating the macro average of F1 score.

\section{Evaluation}
\label{sec:evaluation}
The evaluation of FlaKat aims to answer the research questions listed in Sec.~\ref{sec:problemStatement} and explore the performance of the implementation variations mentioned in Sec.~\ref{sec:workflowAndImplementation}. In Phase II of the workflow, three source code vectorization techniques, doc2vec, code2vec, and tf-idf, were used to vectorize the Java flaky tests after parsing and preprocessing. Dimension reduction techniques, PCA, LDA, Isomap, t-SNE and UMAP, were applied in Phase III to the vector embeddings. The reduced embeddings were analyzed both qualitatively and quantitatively to select the one that best preserves the local and global structure of flaky tests vector embeddings. Evaluation of the machine learning classifier used in Phase IV (KNN, SVM and Random Forest) was based on two metrics: \(F_1\) score and FDC. 

\subsection{Dataset}
\label{sec:dataset}

\begin{figure}[htb]
\begin{tikzpicture}[pie-label/.style = {text width=3cm, align=left, anchor=north west, inner xsep=0pt}]

\def\printonlylargeenough#1#2{\unless\ifdim#2pt<#1pt\relax
#2\printnumbertrue
\else
\printnumberfalse
\fi}
\newif\ifprintnumber
\pie[color = {
        blue!70,
        red!70, 
        green!70, 
        gray!80,
        violet!70,
        yellow!80,
        brown!80},
    sum = auto,
    text = legend,
    radius= 1.8,
    before number=\printonlylargeenough{50},
    after number=\ifprintnumber\fi
    ]{589/Implementation-Dependent,
    307/Order-Dependent,
    133/Order-Dependent Victim,
    109/Non-Deterministic,
    93/Unknown Dependency,
    15/Order-Dependent Brittle,
    11/{\tikz[baseline=1\baselineskip]\node[pie-label]
        {Non-Deterministic- 
        Order-Dependent};}}

\end{tikzpicture}
\caption{The original distribution of categories of flaky tests in the 
dataset.}
\label{fig:datasetPie}
\end{figure}
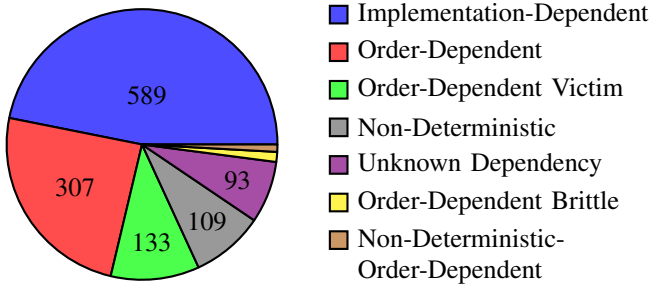

A flaky test dataset that provides their flakiness category labels was required for training the machine learning models and evaluating the prediction. The IDoFT fulfills this requirement and provides thousands of real-world flaky tests that reside in numerous open-source projects. With the corresponding URL of all flaky tests, the FlaKat framework can easily obtain the source code and analyze flaky test cases in Phase I. In total, there were 1,257 flaky tests from 108 open-source projects. The distribution of categories is shown in Fig.~\ref{fig:datasetPie}. Out of all tests, 589 of them were Implementation-Dependent (ID) manifest by~\cite{ref:NonDex}, 307 of them were Order-Dependent (OD) labelled by~~\cite{ref:iDflakies}, 133 and 15 of them were grouped into more specific Order-Dependent Victim (OD-Vic) and Brittle (OD-Brit) by~\cite{ref:iFixFlakies}. In addition, there were 109 Non-Deterministic(NOD) flaky tests, and 93 have Unknown-Dependency (UD) that passes and fails in a test suite or isolation. There were also 11 Non-Deterministic Order-Dependent (NDOD) tests that fail non-deterministically but with significantly different failure rates in different orders~\cite{ref:reproducibilityAndCharacteristics}. Compared to flakyCat~\cite{ref:flakyCat} which uses a small dataset of 343 flaky tests with self-defined labels, IDoFT dataset is much larger and labeled with categories agreed upon within the community thus granting more confidence in the result of FlaKat.

\begin{figure*}[htb]
    \centering
    \includegraphics[width=1\textwidth]{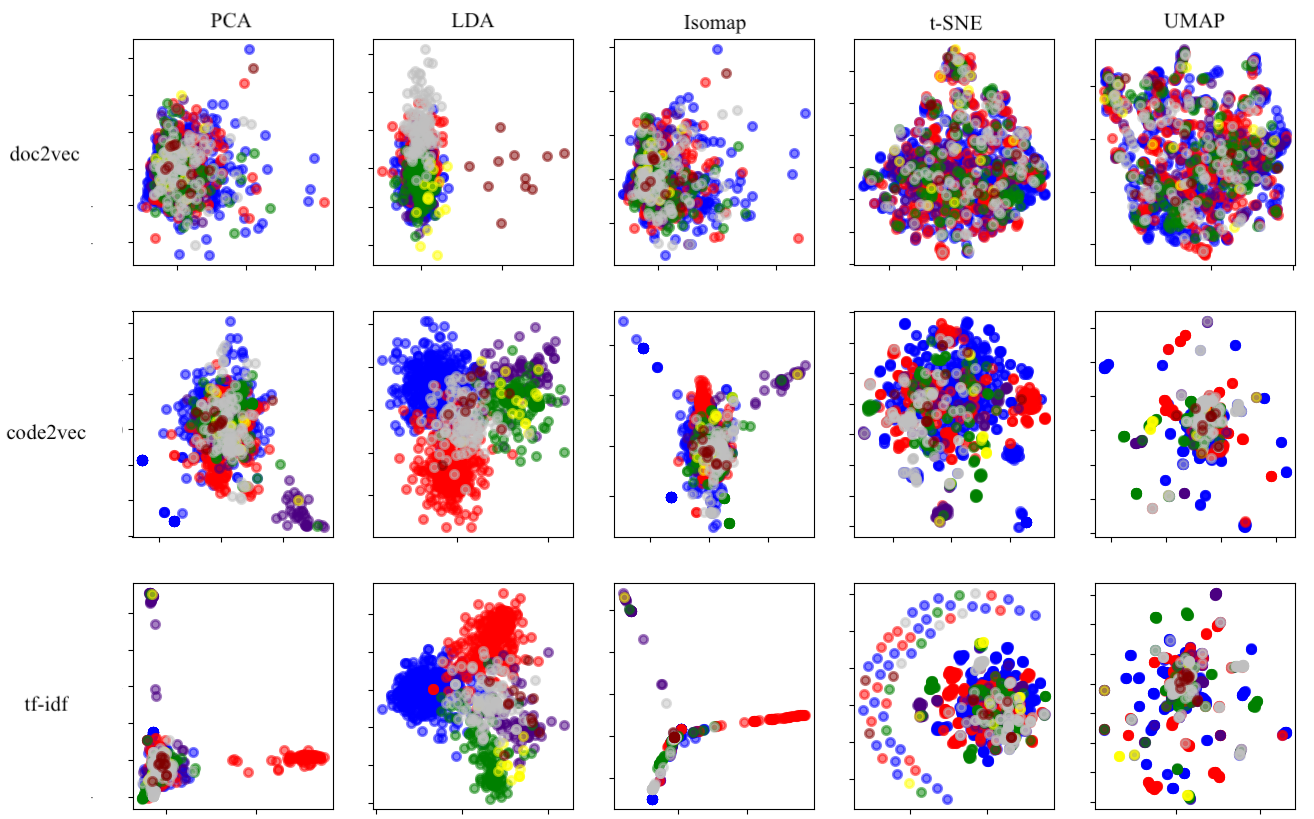}
    \caption{Visualization of data points in reduced vector space.}
    \label{fig:scatterPlots}
\end{figure*}
The imbalance in distribution of flaky test categories could lead to unsatisfactory classification results. Therefore, the sampling techniques mentioned in Phase IV were used to address this challenge. Synthetic data points for minority categories were created via SMOTE. On the other hand, TK was used to remove the bordering majority category data points for better separation between clusters.

\subsection{Vector Embedding Generations}
The parsed raw test cases were then converted into vector embeddings using the techniques mentioned in Phase II of FlaKat in Sec.~\ref{sec:workflowAndImplementation}. The Doc2vec model was trained using the entirety of downloaded test cases. Code2vec used the pre-trained model provided by the author of code2vec~\footnote{https://github.com/tech-srl/code2vec}. Tf-idf did not require any training. The known flaky tests were then converted into vector embedding of 384 features following the default setting of code2vec.

\subsection{Dimensionality Reduction}

Once parsed, the raw test cases were then converted into vector embeddings via doc2vec, code2vec or tf-idf with different dimension reduction techniques applied to them. The techniques considered were PCA, LDA, Isomap, t-SNE, and UMAP. Unlike PCA and LDA, the Isomap, t-SNE, and UMAP reductions all have tunable hyperparameters that affect the result of the dimension reduction. 

\subsubsection{\textbf{Qualitative Analysis}}

The qualitative analysis of the dimensionality reduction techniques was first done by manual inspections of the visualized two-dimensional projection in Fig.~\ref{fig:scatterPlots}. From left to right, the scatter plots are the result of PCA, LDA, Isomap, t-SNE, and UMAP respectively and from top to bottom, the plots were generated using doc2vec, code2vec, and tf-idf respectively.

\noindent{$\bullet$ \textbf{doc2vec:}}
The PCA reduction technique applied on doc2vec embedding did not produce a promising result, as shown in its two-dimensional scatter plot. Most of the data points were located in a single cluster with a few ID (in blue) and OD (in red) data points randomly spread apart. In the plot for LDA, NDOD (in brown) data points were clearly separated from the other categories, which were grouped in a huge cluster along a vertical axis with ND (in grey) data points located at the top, OD (in red) data points in the middle, and the rest at the bottom. The projection of Isomap was quite similar to the one generated using PCA with no obvious clustering between the different categories except a few ID data points randomly located far from a large central cluster. The results from the manifold reduction methods t-SNE and UMAP both appear to be worse compared to the previous techniques.

\noindent{$\bullet$ \textbf{code2vec:}}
In the two-dimensional PCA projection, some UD (in violet) data points were clustered together while the data points from other categories overlap with each other. Compared to the reduced result from doc2vec, this result is marginally better. Reduction by LDA yielded cleaner clustering for all categories except OD-Vic (in green) and OD-Brit (in yellow) data points, which were slightly mixed. Data points in Isomap projections for code2vec embedding were less spread out regardless of the value of neighbours compared to Isomap projection for doc2vec. However, the majority of data points overlap with each other except UD (in violet). Tuning perplexity and the number of iterations for t-SNE generated different reduction results, which yielded small clusters of ID, OD, UD, and OD-Vic while all remaining data points were almost evenly distributed in a central cluster. The results of UMAP for code2vec embedding were also not promising. It exhibited similarly small clusters like t-SNE.

\noindent{$\bullet$ \textbf{tf-idf:}}
As observed in Fig~\ref{fig:scatterPlots}, in the two-dimensional PCA projection, some OD and UD data points were cleanly clustered while the rest were grouped into a single cluster with flaky tests from different categories. In contrast to PCA, LDA yielded decent separation of clusters between all categories, but there was overlapping at the boundaries of clusters, especially for NOD data points. Depending on the number of neighbours, Isomap yielded different two-dimensional projections but the data points were usually densely located along several axes with no obvious clustering. Applying t-SNE reduction also produced different outcomes depending on the perplexity setting and number of iterations but there was always a large cluster with mixed data points from all categories. UMAP should outperform t-SNE~\cite{ref:umap} but the projection showed a huge cluster of flaky tests from different categories mixed together similar to t-SNE.

\begin{tcolorbox}[width=\linewidth, sharp corners=all, colback=white!95!black]

\textbf{Answer to RQ1:} Yes, the observations from some of the two-dimensional projection of vector embeddings show clear separation of flaky test data points from different categories under certain settings. This is \textbf{most notable in the plots from \textbf{code2vec and tf-idf reduced by LDA.}}

\end{tcolorbox}

\definecolor{LightGray}{gray}{0.9}
\begin{table*}[h!]
  \centering
  \renewcommand{\arraystretch}{1.2}
  \begin{tabular}{|p{0.4cm}|c c c c c|c c c c c|c c c c c|}
    \hline
    \multirow{2}{0.4cm}{\textbf{K}} & \multicolumn{5}{c|}{\textbf{doc2vec}} & \multicolumn{5}{c|}{\textbf{code2vec}}& \multicolumn{5}{c|}{\textbf{tf-idf}}\\
    \cline{2-16}
    & \textbf{PCA} & \textbf{LDA} & \textbf{Isomap} & \textbf{t-SNE} & \textbf{UMAP} & \textbf{PCA} & \textbf{LDA} & \textbf{Isomap} & \textbf{t-SNE} & \textbf{UMAP} & \textbf{PCA} & \textbf{LDA} & \textbf{Isomap} & \textbf{t-SNE} & \textbf{UMAP}    \\
    \hline
    
    \rowcolor{LightGray}
    2 & 0.20 & \textbf{0.38} & 0.20 & 0.22 & 0.22 & 0.36 & \textbf{0.50} & 0.43 & 0.54 & 0.44 & 0.40 & 0.53 & 0.45 & \textbf{0.59} & 0.49\\
    
    5 & 0.17 & \textbf{0.37} & 0.17 & 0.19 & 0.19 & 0.34 & \textbf{0.54} & 0.39 & 0.51 & 0.42 & 0.37 & 0.54 & 0.43 & \textbf{0.56} & 0.44\\
    
    \rowcolor{LightGray}
    10 & 0.15 & \textbf{0.39} & 0.16 & 0.17 & 0.20 & 0.34 & \textbf{0.54} & 0.37 & 0.48 & 0.42 & 0.36 & \textbf{0.55} & 0.38 & 0.52 & 0.44\\
    
    20 & 0.15 & \textbf{0.39} & 0.15 & 0.17 & 0.18 & 0.31 & \textbf{0.54} & 0.35 & 0.45 & 0.41 & 0.34 & \textbf{0.56} & 0.35 & 0.49 & 0.41\\
    
    \rowcolor{LightGray}
    50 & 0.14 & \textbf{0.38} & 0.12 & 0.15 & 0.15 & 0.30 & \textbf{0.56} & 0.32 & 0.40 & 0.40 & 0.33 & \textbf{0.59} & 0.33 & 0.41 & 0.36\\
    
    100 & 0.15 & \textbf{0.40} & 0.13 & 0.15 & 0.16 & 0.27 & \textbf{0.56} & 0.28 & 0.37 & 0.37 & 0.30 & \textbf{0.59} & 0.27 & 0.32 & 0.29\\
    
    \rowcolor{LightGray}
    200 & 0.13 & \textbf{0.40} & 0.13 & 0.13 & 0.15 & 0.27 & \textbf{0.56} & 0.24 & 0.33 & 0.32 & 0.30 & \textbf{0.60} & 0.22 & 0.30 & 0.23\\
    
    500 & 0.10 & \textbf{0.39} & 0.14 & 0.12 & 0.12 & 0.26 & \textbf{0.53} & 0.28 & 0.23 & 0.25 & 0.23 & \textbf{0.59} & 0.16 & 0.23 & 0.20\\

    \hline
  \end{tabular}
  \caption{Quantitatively Analysis with doc2vec, code2vec and tf-idf using \(F_1\) score}
  \label{tab:quantitativeAnalysisGeneral}
\end{table*}

\subsubsection{\textbf{Quantitative Analysis}}

The quantitative analysis of dimensionality reduction is done by running the KNN algorithm on the reduced embeddings with various k values. When k (i.e. the number of neighbours to consider) is small, the results represent how well the local structure of a cluster is preserved. On the other hand, when k is large, the results reflect how well data points structures are preserved globally. The minimum possible value for k was 2 due to the nature of the KNN algorithm and the largest value of k was set to 500. In addition, 10-fold cross-validation and sampling were applied for better estimation. The highest \(F_1\) score for each embedding is highlighted for every k value. 

\noindent{$\bullet$ \textbf{doc2vec:}}
In the first column group of Table~\ref{tab:quantitativeAnalysisGeneral}, the \(F_1\) score of applying KNN on the reduced doc2vec embeddings is displayed with the k value ranging from 2 to 500. When k was small, embeddings reduced by LDA displayed much higher performance at 0.38 compared to others, and such superiority was maintained for all chosen k values. When k was large, the performance of LDA peaked at 0.40. The results from PCA, Isomap, t-SNE, and UMAP were similar with \(F_1\) scores of 0.20~0.22 when k was 2, The \(F_1\) scores decreased to 0.13 as k grew larger. Thus, the LDA reduction quantitatively yielded the best-reduced vector embedding both locally and globally for doc2vec.

\noindent{$\bullet$ \textbf{code2vec:}}
The performance of the KNN classifier using the vector embedding generated from code2vec is shown in the second column group of Table~\ref{tab:quantitativeAnalysisGeneral} measured by \(F_1\) score. LDA started at 0.50 when k was at its minimum and increased to 0.56 as k reached 200. It then declined to 0.53 when k was 500. Compared to the results of doc2vec, the LDA reduction for embedding generated by code2vec showed significantly higher performance than the other reduction methods. Additionally, the variance in \(F_1\) score across different k values was also lower for LDA. This makes LDA the best option for code2vec embedding among the considered reduction techniques. The UMAP reduction ranked second, and marginally surpassing PCA, Isomap, and t-SNE in most situations.

\noindent{$\bullet$ \textbf{tf-idf:}}
The outcome of tf-idf embedding was similar to that of code2vec (as shown in the last column of Table~\ref{tab:quantitativeAnalysisGeneral}). When k was at 2, the \(F_1\) score from LDA was at 0.53, which is lower compared to 0.59 obtained for t-SNE. However, the gap between the LDA and t-SNE decreased as k increased, and \(F_1\) score from LDA surpassed t-SNE once k became larger than 10. Although LDA was not the optimal technique for all values of k, its \(F_1\) score was only sub-optimal by a small margin when k was small, while it yielded significantly higher \(F_1\) scores when k was large; thus, we conclude LDA to be the best reduction technique for tf-idf.

\begin{tcolorbox}[width=\linewidth, sharp corners=all, colback=white!95!black]

\textbf{Answer to RQ2:} The reduced embeddings generated by \textbf{LDA best preserves the local and global structure }according to the performance of KNN algorithm measured in \(F_1\) score.

\end{tcolorbox}


\begin{figure*}[!tbp]
     \centering
     \begin{subfigure}{0.49\textwidth}
         \centering
         \includegraphics[width=\textwidth]{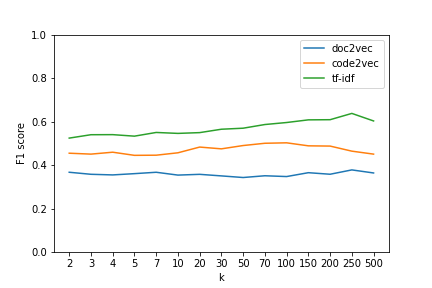}
         \caption{KNN classifier.}
         \label{fig:knn_f1s}
     \end{subfigure}
     \hfill
     \begin{subfigure}{0.49\textwidth}
         \centering
         \includegraphics[width=\textwidth]{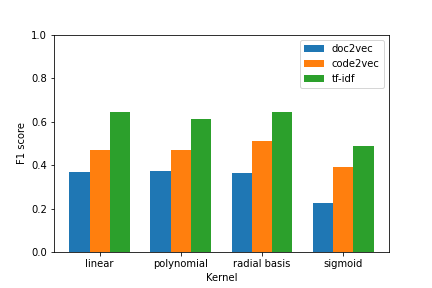}
         \caption{SVM classifier.}
         \label{fig:svm_f1s}
     \end{subfigure}
    
    \caption{Performance of embeddings  on classifiers measured in \(F_1\) score }
    \label{fig:classifier_f1s}
\end{figure*}

\subsection{Prediction Effectiveness}

With the knowledge that vector embeddings of flaky tests are best preserved using LDA dimensionality reduction, different machine learning models were evaluated to process the reduced dimensional vector representation of flaky tests and make classification of their categories. The performance of classification were measured in \(F_1\) score and FDC as the average of 10-fold cross-validation.

\subsubsection{\textbf{\(F_1\) score}}

The results from the KNN classifier for all 3 vector embeddings are shown in Fig.~\ref{fig:knn_f1s}.  It was palpable that predictions based on the embedding generated by doc2vec do not have high performance overall. As the value of k increased, its \(F_1\) score remained around 0.37 regardless of k values. Both code2vec and tf-idf showed improved performance compared to doc2vec, with the latter slightly better than the former across different values of k. When k was at its minimum value of 2, their \(F_1\) scores were at 0.45 and 0.52 respectively. The code2vec approach reached 0.50 when k reached 70 and 100 but could not maintain this performance for higher k values. As k increased, KNN prediction performance using tf-idf also increased until a k value of 500. The highest performance for tf-idf was 0.63, and it occurred when k was at 250. Overall, vector embeddings generated by tf-idf achieved the highest score for KNN.


The performance of different types of kernels, linear, polynomial, radial basis, and sigmoid, were explored for the SVM classifier in combination with the three embeddings. Each kernel type was also paired with the regularization parameter ranging from 0.1 to 100. The highest measured \(F_1\) score is plotted in Fig.~\ref{fig:svm_f1s}. The rank of \(F_1\) scores among doc2vec, code2vec, and tf-idf was the same as the rank from KNN. Here, tf-idf also produced the highest performance and doc2vec produced the lowest. For all 3 embeddings, their \(F_1\) scores were around 0.36, 0.48, and 0.63 respectively across the different kernel types. In addition, their performance deteriorated with the sigmoid kernel and dropped to 0.22, 0.39, and 0.48 respectively. By slight superiority, the highest \(F_1\) score came from tf-idf with linear and radial basis function kernel at 0.64.

\definecolor{LightGray}{gray}{0.9}
\begin{table}[h!]
  \centering
  \renewcommand{\arraystretch}{1.2}
  \begin{tabular}{|p{3.2cm}|c |c |}
    \hline
    \textbf{Parameter} & \textbf{Bound} & \textbf{Typ}e \\
    \hline
    \rowcolor{LightGray}
    max\_depth & 1-200 & Integer \\
    
    min\_impurity\_decrease & 0-0.5 & Float \\
    
    \rowcolor{LightGray}
    min\_samples\_leaf & 1-200 & Integer \\
    
    min\_samples\_split & 2-400 & Integer \\
    
    \rowcolor{LightGray}
    n\_estimators & 100-200 & Integer\\
    
    min\_weight\_fraction\_leaf & 0-0.05 & Float\\
    
    \rowcolor{LightGray}
    max\_leaf\_nodes & 2-400 & Integer\\
    
    criterion & gini, entropy, log\_loss & String \\

    \hline
  \end{tabular}
  \caption{ Bound region of parameters for tuning Random Forest classifier using Bayesian Optimization}
  \label{tab:boundRegion}
\end{table}

\definecolor{LightGray}{gray}{0.9}
\begin{table}[h!]
  \centering
  \renewcommand{\arraystretch}{1.2}
  \begin{tabular}{|p{3.2cm}|c |c |c |}
    \hline
    \textbf{Parameter} & \textbf{doc2vec} & \textbf{code2vec} & \textbf{tf-idf} \\
    \hline
    \rowcolor{LightGray}
    max\_depth & 116 & 30 & 53 \\
    
    min\_impurity\_decrease & 0.008 & 0.004 & 0.020 \\
    
    \rowcolor{LightGray}
    min\_samples\_leaf & 34 & 41 & 70 \\
    
    min\_samples\_split & 217 & 227 & 203 \\
    
    \rowcolor{LightGray}
    n\_estimators & 116 & 30 & 53 \\
    
    min\_weight\_fraction\_leaf & 0.042 & 0.002 & 0.005 \\
    
    \rowcolor{LightGray}
    max\_leaf\_nodes & 286 & 125 & 324 \\
    
    criterion & log\_loss& log\_loss& log\_loss \\
    
    \rowcolor{LightGray}
    \(F_1\) scores & 0.35 & 0.50 & 0.67 \\

    \hline
  \end{tabular}
  \caption{ The optimal value for Random Forest classifier parameters found by Bayesian Optimization}
  \label{tab:optimalParameter}
\end{table}

\begin{table*}[h!]
  \centering
  \renewcommand{\arraystretch}{1.2}
  \begin{tabular}{|p{5cm}|c|c c c|c c c|c c c|}
    \hline
    \multirow{2}{1.5cm}{\textbf{Category}} &\multirow{2}{1.5cm}{\textbf{Percentage of dataset}} & \multicolumn{3}{c|}{\textbf{doc2vec}} & \multicolumn{3}{c|}{\textbf{code2vec}}& \multicolumn{3}{c|}{\textbf{tf-idf}}\\
    \cline{3-11}
    &\!& \textbf{KNN} & \textbf{SVM} & \textbf{RF} & \textbf{KNN} & \textbf{SVM} & \textbf{RF} & \textbf{KNN} & \textbf{SVM} & \textbf{RF}\\
    \hline
    \rowcolor{LightGray}
    Implementation-Dependent & 46.8\% & 0.71 & 0.33 & 0.36       & 0.88 & 0.87 & 0.90    & 0.93 & 0.93 & 0.94 \\
    
    Order-Dependent & 24.4\% & 0.26 & 0.41 & 0.15       & 0.84 & 0.85 & 0.87    & 0.91 & 0.91 & 0.90\\
    
    \rowcolor{LightGray}
    Order-Dependent Victim & 10.5\% & 0.16 & 0.29 & 0.30   & 0.60 & 0.63 & 0.68   & 0.79 & 0.74 & 0.81\\
    
    Non-Deterministic& 8.6\%  & 0.28 & 0.24 & 0.61      & 0.34 & 0.34 & 0.45   & 0.46 & 0.48 & 0.54 \\
    
    \rowcolor{LightGray}
    Unknown Dependency & 7.3\% & 0.12 & 0.19 & 0.19       & 0.49 & 0.49 & 0.56   & 0.60 & 0.57 & 0.56\\
    
    Order-Dependent Brittle & 1.2\% & 0.08 & 0.12 & 0.13  & 0.19 & 0.20 & 0.35   & 0.47 & 0.52 & 0.50 \\
    
    \rowcolor{LightGray}
    Non-Deterministic- 
        Order-Dependent & 0.8\% & 1.0 & 1.0 & 1.0        & 0.14 & 0.15 & 0.21    & 0.33 & 0.38 & 0.42\\

    \hline
    \textbf{Average} & N/A & \textbf{0.37} & \textbf{0.37} & \textbf{0.39}  & \textbf{0.50} & \textbf{0.51} & \textbf{0.59}   & \textbf{0.63} & \textbf{0.64} & \textbf{0.67}\\

    \hline
  \end{tabular}
  \caption{ The \(F_1\) scores of specific category of flaky tests using the optimal configurations for classifiers using embeddings generated by doc2vec, code2vec and tf-idf}
  \label{tab:categoricalF1score}
\end{table*}

Tuning Random Forest classifier for optimal result was more challenging compared to KNN and SVM due to the high number of hyperparameters. Bayesian Optimization mention in Sec.~\ref{sec:workflowAndImplementation} was incorporated into FlaKat with some additional modification to address this issue. The objective function that needed to be maximized were the macro average \(F_1\) score of flaky test prediction using Random Forest classifier on the embedding reduced by LDA through 5-fold cross-validation. Several parameters that impact prediction performance for the Random Forest classifier were chosen for tuning. Their bounded parameter space regions are listed in Table.~\ref{tab:boundRegion} and Table.~\ref{tab:optimalParameter} displays the optimal configurations after 500 iterations. The macro average \(F_1\) score from Random Forest using these setting were 0.35, 0.50 and 0.67 for doc2vec, code2vec, and tf-idf respectively.

From the experiments of applying KNN, SVM and Random Forest on the reduced vector embeddings of doc2vec, code2vec, and tf-idf, a couple of observations can be made. First, there was a large discrepancy in performance between the result of doc2vec compared to the other embedding techniques with every machine learning model used. Second, tf-idf produced slightly higher \(F_1\) scores than the ones produced by code2vec and outperformed doc2vec significantly. Third, the best prediction came from Random Forest classifier using the vector embedding generated by tf-idf.

\begin{tcolorbox}[width=\linewidth, sharp corners=all, colback=white!95!black]

\textbf{Answer to RQ3:} Compared to doc2vec and code2vec, predictions made based on \textbf{tf-idf embedding usually yielded better results}. \textbf{The Random Forest classifier with parameters tuned after Bayesian Optimization obtained the highest \(F_1\) score of 0.67} while SVM with linear kernel and regularization value of 0.1 scored 0.64 and KNN with k set to 250 scored 0.63.

\end{tcolorbox}

\subsubsection{\textbf{Category-specific Results}}

\begin{figure*}[h]
     \centering
     \begin{subfigure}[b]{0.49\textwidth}
         \centering
         \includegraphics[width=\textwidth]{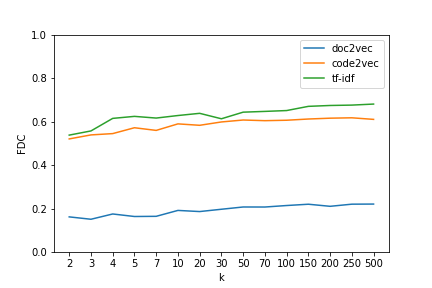}
         \caption{KNN classifier.}
         \label{fig:knn_fdc}
     \end{subfigure}
     \hfill
     \begin{subfigure}[b]{0.49\textwidth}
         \centering
         \includegraphics[width=\textwidth]{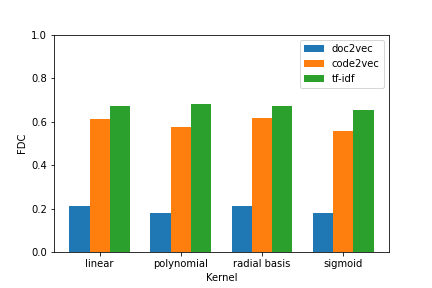}
         \caption{SVM classifier.}
         \label{fig:svm_fdc}
     \end{subfigure}
    
    \caption{Performance of embedding on classifiers measured in FDC }
    \label{fig:classifier_fdc}
\end{figure*}

Looking into the \(F_1\) scores of individual categories revealed more details regarding the effectiveness of flaky test categorization for each classifier on different embeddings. The results displayed in Table.~\ref{tab:categoricalF1score} are from the optimal configuration of each classifiers measured in average \(F_1\) score. The categories are ordered according to their percentage in the original dataset before sampling. For doc2vec embedding, NDOD flaky tests obtain \(F_1\) score of 1 for all KNN, SVM, and Random Forest classifier but its overall averages were still much lower than code2vec and tf-idf. The score is consistent and reproducible but is likely caused by an insufficient amount of NDOD tests before over-sampling. There are more similarities between results from code2vec and tf-idf embeddings. ID and OD typed flaky tests can be predicted with high \(F_1\) score for code2vec embedding (0.90, 0.88, and 0.87) and tf-idf embedding (0.94, 0.93, and 0.93). OD-Brit and NDOD flaky tests show the lowest performance compared to all other categories. For code2vec embedding, their \(F_1\) scores were around 0.15. For tf-idf embedding, their \(F_1\) scores were around 0.38. These results are also aligned with the original distribution of the flaky tests in the dataset before sampling. ID and OD are the most common type of flaky tests and displayed clear clusters in the 2d projection while OD-Brit and NDOD, being the rarest, were mixed with other flaky tests from OD-Vic, NOD, and UD.

\subsubsection{\textbf{Flakiness Detection Capacity}}

A well-established weighting policy for machine learning models has yet to be found and is beyond the scope of this paper. Thus, the Flakiness Detection Capacity (FDC) was proposed to better compare the final results form the classifiers using macro or inversely weighted \(F_1\) score. It is a new metric based on intrusion detection capacity derived from the field of information-theoretic analysis as formulated in Appendix~\ref{app:fdc}. To show the advantages of FDC as a metric for flakiness categorization performance, we demonstrate its high degree of consistency and discriminancy relative to the \(F_1\) score~\cite{ref:auc}. The details illustrating these advantages can be found in Appendix A.

To calculate consistency index C with a decently large size of samples using Eq.~\ref{equ:consistency}, KNN, SVM, and Random Forest classifiers were trained and tested using the balanced IDoFT dataset with 5-fold splits and shuffling. The result of FDC and \(F_1\) score were then recorded and compared. The process was repeated 5, 10, 20, and 50 times for consistency. The value of C remained stable at 0.79, 0.79, and 0.77 for KNN, SVM, and Random Forest classifier,  which was higher than the required 0.5 as stated in the definition of the consistency index.  

A similar procedure was applied on calculating discriminancy index D from Eq.~\ref{equ:discriminancy}. For KNN, SVM, and Random Forest classifier, their corresponding D values were 1.96, 1.93, and 1.86, which are all higher than 1; therefore, FDC is also more discriminant than \(F_1\) score.

The performance measured in FDC for KNN is displayed in Fig.~\ref{fig:knn_fdc}. The ranking of the three vector embeddings was the same as the ranking measured in \(F_1\) score. However, it does bring more insight regarding the impact of changes to the value of k. Unlike the \(F_1\) score, FDC for doc2vec were significantly lower then the other two. The gap between cod2vec and tf-idf was also smaller. FDC of doc2vec started at 0.16 (when k was small) and then gradually reached 0.22 as k increased to 500. A similar trend appeared in both code2vec and tf-idf results, where their FDCs were 0.52 and 0.53 respectively when k equaled to 2, and 0.61 and 0.68 respectively when k equaled to 500. For all k values, tf-idf embeddings produced a higher FDC value than the others.

In Fig.~\ref{fig:svm_fdc}, the FDC measurements of SVM on the embeddings were almost identical to their \(F_1\) score plot. Regardless of kernel type, the output from tf-idf was always more correlated to its vector embedding compared to code2vec. The optimal kernel type under FDC was the polynomial kernel with a score of 0.68. Linear and radial basis function kernel ranked second together with slightly lower FDCs at 0.67. This observation is different from the earlier scenarios where performance were measured in \(F_1\) score. Furthermore, doc2vec vector embeddings did not yield promising results. 

\begin{table}[h!]
  \centering
  \renewcommand{\arraystretch}{1.2}
  \begin{tabular}{|p{3.2cm}|c |c |c |}
    \hline
    \textbf{Parameter} & \textbf{doc2vec} & \textbf{code2vec} & \textbf{tf-idf} \\
    \hline
    \rowcolor{LightGray}
    max\_depth & 142 & 140 & 23 \\
    
    min\_impurity\_decrease & 0.018 & 0.073 & 0.005 \\
    
    \rowcolor{LightGray}
    min\_samples\_leaf & 66 & 16 & 105 \\
    
    min\_samples\_split & 47 & 76 & 66 \\
    
    \rowcolor{LightGray}
    n\_estimators & 113 & 149 & 104 \\
    
    min\_weight\_fraction\_leaf & 0.039 & 0.033 & 0.002 \\
    
    \rowcolor{LightGray}
    max\_leaf\_nodes & 148 & 226 & 338 \\
    
    criterion & entropy & log loss & gini \\
    
    \rowcolor{LightGray}
    FDC & 0.22 & 0.65 & 0.70 \\

    \hline
  \end{tabular}
  \caption{ The optimal value for Random Forest classifier parameters found Bayesian Optimization with FDC as objective function}
  \label{tab:optimalParameterFDC}
\end{table}

The result of Bayesian Optimaiztion using FDC as objective function is shown in Table.~\ref{tab:optimalParameterFDC}. Compared to the optimal configuration collected earlier, the classifier using doc2vec embedding obtained higher minimum impurity required for splitting nodes and lower maximum number of leaf nodes. This indicated the splits were less common and classifier tuned in \(F_1\) score likely suffered over-fitting. Code2vec embedding yielded the opposite outcome, since the max depth of the tree and max number of leaf nodes increased while the minimum sample leaf size and minimum sample required for splitting decreased. The overall splitting was more fine-grained compared to the previous experiment. The difference between the results from tf-idf embedding were relatively smaller. The maximum depth decreased while the minimum impurity required for splitting also decreased. Their FDCs were 0.22, 0.65, and 0.70 respectively, which is consistent with the results using \(F_1\) score as the objective function.

\begin{tcolorbox}[width=\linewidth, sharp corners=all, colback=white!95!black]

\textbf{Answer to RQ4:} Yes, FDC is a better metric for evaluating flaky test categorization due to its higher consistency and discriminancy compared to \(F_1\) score. Among all combinations of vector embeddings and classifiers, the highest FDC was obtained from \textbf{tf-idf embedding and Random Forest classifier with parameters tuned by Bayesian Optimization}.

\end{tcolorbox}

\section{Related Work}
\label{sec:relatedWork}

Test failures due to flakiness are not rare events and have a significant impact on the efficiency and reliability of software testing~\cite{ref:SurveyOfFlaky}. A study conducted at Google found that 41\% of the tests that alternate between pass and failure at least once were flaky~\cite{ref:googleScaleTesting}. Another study at Microsoft identified flaky test failures within 26\% of the total sampled builds~\cite{ref:rootCausingFlaky}.  In addition, 5.7\% of all the failed builds from 80 million test runs in 30 days were caused by a small fraction (0.02\%) of flaky test failures~\cite{ref:lifecycleOfFlakyTest}. Simply ignoring flaky tests could have serious consequences. The investigation into Firefox crash reports suggests that ignoring test failures, even if they are flaky, can lead to a higher incidence of crashes~\cite{ref:impactFlakyTestOnCrashFirefox}. These results indicate that flaky tests can limit the efficiency of continuous integration by causing build failures that require time-consuming manual intervention from software developers.

The most straightforward techniques for automatically detecting flaky tests are based on repeatedly re-running them~\cite{ref:deFlaker,ref:reproducibilityAndCharacteristics}. DeFlaker~\cite{ref:deFlaker} analyzes the difference in code coverage between consecutive versions of the same software with unstable outcomes. Spectrum-based fault localization~\cite{ref:spectrumBasedFaultLocalization} improved on the traditional coverage-based approach and can narrow down flaky tests in Python projects. One study presented iDFlakies, which is a framework based on executing test suites in randomized orders to detect order-dependent flaky tests~\cite{ref:iDflakies} and ~\cite{ref:detectCharacerizeTame}. NonDex~\cite{ref:NonDex} is an approach that randomizes the implementations of various Java classes with non-deterministic specifications to identify implementation-dependent flaky tests~\cite{ref:assumptionOfNonDeterSpec}. Another study presented Shaker, which targets flaky tests of the asynchronous wait and concurrency categories by introducing CPU and memory tasks to affect the ordering of regression tests~\cite{ref:shaker}. Techniques from the field of machine learning have also been applied to flaky test detection, such as FlakeFlagger~\cite{ref:flakeFlagger} and FLAST~\cite{ref:FLAST}. These studies considered the presence of particular keywords in test case source code or other general test characteristics, such as execution time and line of code, as potential predictors of flaky tests~\cite{ref:vocabOfFlaky,ref:replicationVocabOfFlaky}. Classifier trained on the flip rate, change of files and size of latest pull request also proved to be efficient in predicting flakiness~\cite{ref:predicionUsingEvolutionHistory}. A new language model-based predictor for flaky test called Flakify statically predicts flakiness without accessing the production code in a black-box manner~\cite{ref:Flakify}. A new feature set, Flake16~\cite{ref:flake16} is developed to help improve ML-based detection for order-dependent and non-order-dependent falky test in Python projects.

Depending on the cause of flakiness, the options for mitigating and repairing flaky tests may differ. In general, consistent coding guidelines and stable infrastructures are crucial foundation for avoiding flaky tests~\cite{ref:qualitativeStudyOnSourcesImpactsMitigation}. Based on the findings from historical commits in Apache Software Foundation projects and surveying Mozilla developers, two studies~\cite{ref:DeveloperPerspective,ref:empiricalAnalysisFlakyTest2014} concluded that the most common type of fix for flaky tests of the asynchronous wait category involved a \texttt{waitFor} method or its equivalent. As for order-dependent tests that fail when run in isolation but pass when run with some other test~\cite{ref:iFixFlakies}, the most common fixes are to eliminate the dependency between its victim and brittle in the allocation and de-allocation of test methods by techniques such as creating a duplicate instance of some shared resource~\cite{ref:DeveloperPerspective,ref:empiricalAnalysisFlakyTest2014}. The iFixFlakies framework~\cite{ref:iFixFlakies} was presented as the automatic repair of order-dependent flaky tests, which uses program statements from elsewhere in the test suite. Another study presented a template-based repair technique called DexFix~\cite{ref:DexFix,ref:assumptionOfNonDeterSpec} for automatically repairing implementation-dependent flaky tests, such as those detected by the previously mentioned detection tool NonDex~\cite{ref:NonDex}. Other than execution order, code instrumentation could also lead to test flakiness but are very rare in practice~\cite{ref:effectOfInstrumentation}.

\section{Threats to Validity}
\label{sec:threatToValidity}

There are several threats to the validity of this study. Some follow from the design decisions while others follow from the nature of the data used.

\noindent{\textbf{External-Size and labels of the dataset:} Though the findings from the evaluations are supported by carefully formulated experiments, better results are possible if a larger dataset was available with more precise labels that are closely related to the content of the test case instead of its behaviour. One phenomenon observed in the dimensionality reduction of both tf-idf and code2vec vector representations was that there was a big cluster consisting of all categories of flaky tests for the two-dimensional projection from t-SNE and UMAP.

\noindent{\textbf{Internal-Overfitting:} Some reduced two-dimensional projections did not provide a meaningful result, especially ones from doc2vec. Certain parts of the data do not show clean separation between categories in a huge cluster, and the optimal prediction configuration happened when the cleanly separated clusters were correctly predicted and the mixed overlapping clusters were rigorously split. More effort is required to handle the overlapping data points.}

\section{Conclusion \& Future Work}
\label{sec:conclusion}

This research presents the FlaKat framework for efficient, static flaky test categorization. First, the motivation behind building a novel flaky test categorization framework using machine-learning approach was presented. Then, the workflow and details of the implementation of the framework with doc2vec, code2vec, and tf-idf source code representation were illustrated. The final evaluation was done with real-world data from the IDoFT dataset via two metrics \(F_1\) score and FDC. The results illustrated that both code2vec and tf-idf embeddings can closely reflect the flakiness category of test cases and help machine learning classifiers yield accurate predictions on certain categories of flakiness. The existing framework on flaky test categorization~\cite{ref:flakyCat} focused on the difference in keywords instead of the root cause. More studies are required to determine which set of labels is best for various downstream tasks.

\appendix[A. Flakiness Detection Capacity]
\label{app:fdc}
Flakiness Detection Capacity (FDC) is the ratio between the mutual information of vectorized embedding input and its category output to the entropy of input and can be calculated with the equation below.
\begin{equation}
    \begin{aligned}
        FDC & = \dfrac{I(c_{in};c_{out})}{H(c_{in})}
        \\
        & = \dfrac{\sum_{c_{in}}\sum_{c_{out}}p(c_{in},c_{out})\,log\dfrac{p(c_{in},c_{out})}{p(c_{in})p(c_{out})}}{-\sum_{c_{in}}p(c_{in})\,log\,p(c_{in})}
    \end{aligned}
    \label{equ:fdc}
\end{equation}
where \(c_{in}\) and \(c_{out}\) are the actual and predicted categories of flaky tests at the input and output of the machine learning classifier. Its value is the result of the division of the \(mutual\,\,information\,I\) between Input Category (\(c_{in}\)) and Output Category (\(c_{out}\)) over the \(entropy\,\,H\) of Input Category (\(c_{in}\)). The value of \(I\) and \(H\) can be calculated with marginal probability mass function \(p(c_{in})\) and \(p(c_{out})\) and joint probability mass function  \(p(c_{in},c_{out})\).

To prove one metric \textit{strictly} superior than another is difficult and unlikely. However, showing FDC is relatively more consistent and discriminant than \(F_1\) score is sufficient to showcase its value.

\textbf{Consistency:} defined in~\cite{ref:auc} stated that for two measures f, g on domain \(\Psi\) which contains numerous sets of flaky tests, let \(R = \{(a,b)|a,b \in \Psi, f(a)>f(b), g(a)>g(b)\}, S = \{(a,b)|a,b \in \Psi, f(a)>f(b), g(a)<g(b)\} \) the degree of consistency of f and g is C \((0<C<1)\). 
 
\begin{equation}
    \begin{aligned}
        C & = \dfrac{|R|}{|R|+|S|}
    \end{aligned}
    \label{equ:consistency}
\end{equation}
Given the definition, FDC was more consistent than \(F_1\) score when C was larger than 0.5. It means among all combinations of set \(a\) and set \(b\), the occurrence where FDC and \(F_1\) score agree with each other, \(FDC(a)>FDC(b) \) AND \( F_1 score(a)>F_1 score(b)\), was more frequent than occurrences where they disagree.

\textbf{Discriminancy:} defined in ~\cite{ref:auc} stated that for two measures f, g on domain \(\Psi\) , let \(P = \{(a,b)|a,b \in \Psi, f(a)\neq f(b), g(a)=g(b)\}, Q = \{(a,b)|a,b \in \Psi, g(a)=g(b), f(a)\neq f(b)\} \) the degree of discriminancy of f and g is D \((D>0)\), where 
 
\begin{equation}
    \begin{aligned}
        D & = \dfrac{|P|}{|Q|}
    \end{aligned}
    \label{equ:discriminancy}
\end{equation}

Given the definition, FDC was more discriminant than \(F_1\) score when D was larger than 1. It means among all combinations of set \(a\) and set \(b\), the occurrence of FDC producing different results while \(F_1\) score produced same results, \(FDC(a)\neq FDC(b) \) AND \( F_1 score(a)=F_1 score(b)\), was more frequent than the contrary scenario.


\begin{thebibliography}{00}

\bibitem{ref:whenWeTalkFlakiness} M. Barboni, A. Bertolino, G. De Angelis, “What we talk about when we talk about software test flakiness", \textit{Quality of Information and Communications Technology (QUATIC)}, vol 1439, pp. 29-39, 2021.

\bibitem{ref:faultLocalization}B. Vancsics, T. Gergely, and A. Beszédes, “Simulating the effect of test flakiness on fault localization effectiveness", \textit{International Workshop on Validation, Analysis and Evolution of Software Tests (VST)}, pp. 28–35, 2020.


\bibitem{ref:mutationTesting}A. Shi, J. Bell, and D. Marinov, “Mitigating the effects of flaky tests on mutation testing", \textit{International Symposium on Software Testing and Analysis (ISSTA)}, pp. 296–306, 2019.

\bibitem{ref:DeveloperPerspective}M. Eck, F. Palomba, M. Castelluccio, and A. Bacchelli, “Understanding flaky tests: The developer’s perspective", \textit{Joint Meeting of the European Software Engineering Conference and the Symposium on the Foundations of Software Engineering (ESEC/FSE)}, pp. 830–840, 2019.

\bibitem{ref:killingGatekeeper} F. Lacoste,  "Killing the gatekeeper: Introducing a continuous integration system", \textit{Agile Conference (AGILE)}, pp. 387–392, 2009.

\bibitem{ref:largeScaleStudy} W. Lam, S. Winter, A. Wei, T. Xie, D. Marinov, and J. Bell. , “A large-scale longitudinal study of flaky tests", \textit{ACM on Programming Language,} vol. 4, pp. 1-29, 2020.

\bibitem{ref:flakeFlagger} A. Alshammari, C. Morris, M. Hilton, and J. Bell, “FlakeFlagger: Predicting flakiness without rerunning tests", \textit{International Conference on Software Engineering (ICSE)}, pp. 1572-1584, 2021.

\bibitem{ref:modelRankAtApple} E. Kowalczyk, K. Nair, Z. Gao, L. Silberstein, T. Long, and A. Memon, “Modeling and ranking flaky tests at Apple", \textit{International Conference on Software Engineering: Software Engineering in Practice (ICSE-SEIP)}, pp. 110–119, 2020.


\bibitem{ref:eventOrder} Z. Dong, A. Tiwari, X. L. Yu, and A. Roychoudhury, “Flaky test detection in Android via event order exploration", \textit{ACM Joint Meeting on European Software Engineering Conference and Symposium on the Foundations of Software Engineering (ESEC/FSE)}, pp. 367–378, 2021.

\bibitem{ref:iDflakies} W. Lam, R. Oei, A. Shi, D. Marinov, and T. Xie, “IDFlakies: A framework for detecting and partially classifying flaky tests", \textit{International Conference on Software Testing, Verification and Validation (ICST)}. pp. 312–32, 2019.

\bibitem{ref:assosiationRule} K. Herzig and N. Nagappan, “Empirically detecting false test alarms using association rules", \textit{International Conference on Software Engineering (ICSE)}, vol. 2, pp. 39–48, 2015.

\bibitem{ref:BayesianNetwork} T. M. King, D. Santiago, J. Phillips and P. J. Clarke, “Towards a Bayesian Network Model for Predicting Flaky Automated Tests," \textit{International Conference on Software Quality, Reliability and Security Companion (QRS-C)}, pp. 100-107, 2018.



\bibitem{ref:vocabOfFlaky} G. Pinto, B. Miranda, S. Dissanayake, M. D. Amorim, C. Treude, A. Bertolino, and M. D’amorim, “What is the vocabulary of flaky tests?", \textit{International Conference on Mining Software Repositories (MSR)}, pp. 492–502, 2020.

\bibitem{ref:empiricalAnalysisFlakyTest2014} Q. Luo, F. Hariri, L. Eloussi, and D. Marinov, “An empirical analysis of flaky tests", \textit{Symposium on the Foundations of Software Engineering (FSE)}, pp. 643–653, 2014.


\bibitem{ref:iFixFlakies} A. Shi, W. Lam, R. Oei, T. Xie, and D. Marinov, “iFixFlakies: A framework for automatically fixing order- dependent flaky tests", \textit{Joint Meeting on European Software Engineering Conference and Symposium on the Foundations of Software Engineering (ESEC/FSE)}, pp. 545–555, 2019.


\bibitem{ref:DexFix} P. Zhang, Y. Jiang, A. Wei, V. Stodden, D. Marinov, and A. Shi, “Domain-specific fixes for flaky tests with wrong assumptions on underdetermined specifications", \textit{International Conference on Software Engineering (ICSE)}, pp. 50–61, 2021.


\bibitem{ref:flakyFix} S. Fatima, H. Hemmati, and L. Briand, “FlakyFix: Using Large Language Models for Predicting Flaky Test Fix Categories and Test Code Repair", \textit{digital preprint, arXiv:2307.00012}, 2024.

\bibitem{ref:FLAST} A. Bertolino, E. Cruciani, B. Miranda, and R. Verdecchia, “Know your neighbor: Fast static prediction of test flakiness", \textit{IEEE Access}, vol. 9, pp. 76119-76134, 2021.


\bibitem{ref:doc2vec} Q. V. Le, and T. Mikolov. “Distributed Representations of Sentences and Documents”, \textit{International Conference on Machine Learning (ICML)}, pp. 1188-1196, 2014.

\bibitem{ref:code2vec} U. Alon, M. Zilberstein, O. Levy and E. Yahav,  “code2vec: learning distributed representations of code”, \textit{ACM on Programming Languages}, pp. 1-29, 2019.

\bibitem{ref:tfidf} K. S. Jones, "A Statistical Interpretation of Term Specificity and its Application in Retrieval", \textit{Journal of Documentation}, vol. 28, no. 1, pp. 11-21, 1972

\bibitem{ref:improvedSoftwareCategorization} A. LeClair, Z. Eberhart and C. McMillan, “Adapting Neural Text Classification for Improved Software Categorization", \textit{International Conference on Software Maintenance and Evolution (ICSME)}, pp. 461-472, 2018.

\bibitem{ref:intrusionDetection} G. Gu, P. Fogla, D. Dagon, W. Lee, and B. Skorić, “Measuring intrusion detection capability: an information-theoretic approach", \textit{ACM Symposium on Information, Computer and Communications Security (ASIACCS)}, pp. 90–101, 2006.


\bibitem{ref:thesis} S. Lin, "FlaKat: A Machine Learning-Based Categorization Framework for Flaky Tests", \textit{UWSpace}, http://hdl.handle.net/10012/19125, 2023.




\bibitem{ref:SurveyOfFlaky}O. Parry, G. M. Kapfhammer, M. Hilton, P. McMinn, “A Survey of flaky tests", \textit{ACM Transactions on Software Engineering Methodology}, 31(1), pp. 1-74, 2022.

\bibitem{ref:googleScaleTesting}A. Memon, Z. Gao, B. Nguyen, S. Dhanda, E. Nickell, R. Siemborski, and J. Micco, “Taming Google-Scale continuous testing", \textit{International Conference on Software Engineering: Software Engineering in Practice (ICSE-SEIP)}, pp. 233–242, 2017.

\bibitem{ref:rootCausingFlaky}W. Lam, P. Godefroid, S. Nath, A. Santhiar, and S. Thummalapenta, “Root causing flaky tests in a large-scale industrial setting", \textit{International Symposium on Software Testing and Analysis (ISSTA)}, pp. 204–215, 2019.

\bibitem{ref:lifecycleOfFlakyTest}W. Lam, K. Muşlu, H. Sajnani, and S. Thummalapenta, “A study on the lifecycle of flaky tests", \textit{International Conference on Software Engineering (ICSE)}, pp. 1471–1482, 2020.


\bibitem{ref:impactFlakyTestOnCrashFirefox}M. T. Rahman and P. C. Rigby, “The impact of failing, flaky, and high failure tests on the number of crash reports associated with Firefox builds", \textit{Joint Meeting on European Software Engineering Conference and Symposium on the Foundations of Software Engineering (ESEC/FSE)}, pp 857–862, 2018.


\bibitem{ref:deFlaker} J. Bell, O. Legunsen, M. Hilton, L. Eloussi, T. Yung, and D. Marinov, “DeFlaker: Automatically detecting flaky tests", \textit{International Conference on Software Engineering (ICSE)}, pp. 433–444, 2018.


\bibitem{ref:spectrumBasedFaultLocalization}M. Gruber, and G. Fraser, “Debugging Flaky Tests using Spectrum-based Fault Localization", \textit{International Conference on Automation of Software Test (AST)}, 2023.


\bibitem{ref:reproducibilityAndCharacteristics} W. Lam, S. Winter, A. Astorga, V. Stodden, and D. Marinov, “Understanding reproducibility and characteristics of flaky tests through test reruns in Java projects", \textit{International Conference on Software Reliability Engineering (ISSRE)}, pp. 403–413, 2020.

\bibitem{ref:detectCharacerizeTame} W. Lam, “Detecting, characterizing, and taming flaky tests",  Ph.D dissertation, Dep. of Computer Science, Univ. of Illinois, Urbana-Champaign, 2021

\bibitem{ref:NonDex} A. Gyori, B. Lambeth, A. Shi, O. Legunsen, and D. Marinov, “NonDex: A tool for detecting and debugging wrong assumptions on Java API Specification", \textit{Symposium on the Foundations of Software Engineering (FSE)}, pp. 223–233, 2015.

\bibitem{ref:effectOfInstrumentation}S. Rasheed, J. Dietrich, and A. Tahir, “On the Effect of Instrumentation on Test Flakiness", \textit{International Conference on Automation of Software Test (AST)}, 2023.


\bibitem{ref:flakyCat} A. Akli, G. Haben, S. Habchi, M. Papadakis and Y. Le Traon, "Predicting Flaky Tests Categories using Few-Shot Learning," \textit{ International Conference on Automation of Software Test (AST)}, 2023.

\bibitem{ref:assumptionOfNonDeterSpec} A. Shi, A. Gyori, O. Legunsen, and D. Marinov, “Detecting assumptions on deterministic implementations of non-deterministic specifications", \textit{International Conference on Software Testing, Verification and Validation (ICST)}. pp. 80–90, 2016.


\bibitem{ref:shaker} D. Silva, L. Teixeira, and M. D’Amorim, “Shake It! Detecting flaky tests caused by concurrency with Shaker", \textit{International Conference on Software Maintenance and Evolution (ICSME)}, pp. 301–311, 2020.

\bibitem{ref:replicationVocabOfFlaky} G. Haben, S. Habchi, M. Papadakis, M. Cordy, and Y. Le Traon, “A replication study on the usability of code vocabulary in predicting flaky tests", \textit{International Conference on Mining Software Repositories (MSR)}, pp. 219-229, 2021.

\bibitem{ref:predicionUsingEvolutionHistory} M. Gruber, M. Heine, N. Oster, M. Philippsen and G. Fraser, "Practical Flaky Test Prediction using Common Code Evolution and Test History Data," \textit{International Conference on Software Testing, Verification and Validation (ICST)}, pp. 210-221, 2023.
\bibitem{ref:Flakify} S. Fatima, T. A. Ghaleb and L. Briand, "Flakify: A Black-Box, Language Model-based Predictor for Flaky Tests", \textit{Transaction on Software Engineering (TSE)}, pp 1-7, 2022.


\bibitem{ref:flake16} O. Parry, G. M. Kapfhammer, M. Hilton and P. McMinn, "Evaluating Features for Machine Learning Detection of Order- and Non-Order-Dependent Flaky Tests," \textit{International Conference on Software Testing, Verification and Validation (ICST)}, pp. 93-104, 2022.



\bibitem{ref:qualitativeStudyOnSourcesImpactsMitigation} S. Habchi, G. Haben, M. Papadakis, M. Cordy, and Y. L. Traon, "A qualitative study on the sources, impacts, and mitigation strategies of flaky tests," \textit{International Conference on Software Testing, Verification and Validation (ICST)}, pp. 244-255, 2022.
\bibitem{ref:bugsInTestCode} A. Vahabzadeh, A. A. Fard, and A. Mesbah, “An empirical study of bugs in test code", \textit{International Conference on Software Maintenance and Evolution (ICSME)}, pp. 101–110, 2015.
 


\bibitem{ref:empiricalRevisitTestIndependence} S. Zhang, D. Jalali, J. Wuttke, K. Muşlu, W. Lam, M. D. Ernst, and D. Notkin, “Empirically revisiting the test independence assumption", \textit{International Symposium on Software Testing and Analysis (ISSTA)}, pp. 385–396, 2014.

\bibitem{ref:flakyTestsInML} S. Dutta, A. Shi, R. Choudhary, Z. Zhang, A. Jain, and S. Misailovic, “Detecting flaky tests in probabilistic and machine learning applications", \textit{International Symposium on Software Testing and Analysis (ISSTA)}, pp. 211–224, 2020.

\bibitem{ref:oracleApproximations} M. Nejadgholi and J. Yang, “A study of oracle approximations in testing deep learning libraries", \textit{International Conference on Automated Software Engineering (ASE)}, pp. 785–796, 2019.










\bibitem{ref:umap} L. McInnes and J. Healy. “UMAP: Uniform Manifold Approximation and Projection for Dimension Reduction”, \textit{Journal of Open Source Software}, 2018.

\bibitem{ref:auc} C.X. Ling, J. Huang, H. Zhang, "AUC: A Better Measure than Accuracy in Comparing Learning Algorithms", \textit{ Advances in Artificial Intelligence}, vol. 2671, pp. 329–341.
2003
\bibitem{ref:StoneDetector} W. Amme, T. S. Heinze and A. Schäfer, “You Look so Different: Finding Structural Clones and Subclones in Java Source Code", \textit{International Conference on Software Maintenance and Evolution (ICSME)}, pp. 70-80, 2021.

\bibitem{ref:CCLearner} L. Li, H. Feng, W. Zhuang, N. Meng and B. Ryder, “CCLearner: A Deep Learning-Based Clone Detection Approach", \textit{International Conference on Software Maintenance and Evolution (ICSME)}, pp. 249-260, 2017.


\end{thebibliography}
\end{document}